\documentclass[10pt,onecolumn]{myart}
\usepackage{graphics}
\usepackage{amssymb}
\newcommand{\vek}[1]{\mbox{\bf #1}}

\begin{document}
\begin{frontmatter}

\title{A photon mass on the brane}

\author{Rainer Dick}, 
\ead{rainer@sask.usask.ca}
\author{Dana M. E. McArthur}

\address{Department of Physics and Engineering Physics,
 University of Saskatchewan,\\ 116 Science Place,
 Saskatoon, SK S7N 5E2, Canada}

\begin{abstract}
We discuss the impact of a bulk photon mass in a
Dvali--Gabadadze--Porrati type brane model with Maxwell
terms both on the brane and in the bulk, as proposed
by Dvali, Gabadadze and Shifman.

The motivation to include the bulk photon mass is to suppress
radiation loss into the bulk.

We point out that this modifies the photon propagator
in such a way that it generates a small photon mass on the
brane. Compatibility with present
bounds on a photon mass imply that the transition
to five-dimensional distance laws for the
electromagnetic potentials
would appear only at super-horizon length scales, thus excluding any
direct detection possibility of a transition from
four-dimensional to five-dimensional distance laws
in electromagnetic interactions.

We also include results on fermion propagators with Dirac
terms on the brane and in the bulk.

\end{abstract}
\begin{keyword}
 brane worlds \sep DGP brane model 
\PACS 04.50.+h 
\end{keyword}
\end{frontmatter}

\section{Introduction}\label{intro}

In the present letter we would like to report results
on the free propagators of fermions and gauge fields 
in a brane model where penetration and radiation into
an infinitely large transverse dimension is suppressed by bulk
mass terms.

After the pioneering work of Dvali, Gabadadze and Porrati
on the emergence of an interpolating gravitational potential
between four and five dimensions in a model with Einstein--Hilbert
terms both on a brane and in the bulk \cite{DGP}, Dvali, Gabadadze and
Shifman have shown that a similar interpolating static potential
exists for gauge fields with both brane and bulk kinetic terms \cite{DGS}.

A generic feature of these models is the prediction that the
four-dimensional potentials are realized at length scales
well below a transition scale $\ell\simeq\mathcal{M}^{-1}$ 
beyond which the five-dimensional
distance laws become dominant. In Gaussian normal coordinates
around a 3-brane $x^\perp=0$ the relevant Lagrangians can be combined
into\footnote{
Our conventions follow \cite{rd1}.
In particular, the perpendicular coordinate $x^\perp$ is chosen such 
that $|x^\perp|$ is the proper or geodesic distance from the 3-brane 
$\mathcal{M}_3$
of the DGP model, whereas the first 4 coordinates $x^\mu=\{t,\vek{x}\}$
cover patches of constant proper distance from $\mathcal{M}_3$:
\[
ds^2=g_{\mu\nu}dx^\mu dx^\nu+(dx^{\perp})^2.
\]
The reduced bulk and brane Planck masses are $m_4$ and $m_3$,
and the bulk and brane gauge couplings are $q_4$ and $q_3$,
respectively.

 The extrinsic curvature tensor in these coordinates is 
\[
K_{\mu\nu}=-\frac{1}{2}\partial_{\perp}g_{\mu\nu}.
\]
}
\begin{eqnarray}\label{eq:actionDG}
S&=&\int dt\int d^{3}\vek{x}\int dx^\perp\,\sqrt{-g}
\left(\frac{m_4^3}{2}R 
-\frac{1}{4q_4^2}F_{MN}{}^aF^{MN}{}_a\right)\\ \nonumber
&{}&
+\int dt\int d^{3}\vek{x}\,\sqrt{-g}\left(
\frac{m_{3}^{2}}{2}R^{(3)}
\left.
-m_4^{3}\overline{K}
-\frac{1}{4q_3^2}F_{\mu\nu}{}^aF^{\mu\nu}{}_a
+\mathcal{L}\right)\right|_{x^\perp=0},
\end{eqnarray}
where $\overline{K}$ is the mean extrinsic curvature scalar
on the brane\footnote{See \cite{CR} for the first realization
that brane models should include a 
Gibbons--Hawking term, and \cite{DM,rda} for a discussion
of alternatives to a Gibbons--Hawking term. See also \cite{lalak}
for a further discussion.} \cite{rd1}:
\begin{equation}\label{eq:meanK}
\overline{K}=\frac{1}{2}
\lim_{\epsilon\to +0}\left[K|_{x^\perp =-\epsilon}
+K|_{x^\perp =\epsilon}\right].
\end{equation}
$R^{(3)}$ is the intrinsic curvature scalar on the brane, and
the Lagrangian $\mathcal{L}$ accounts for 
scalar and fermionic matter degrees of freedom on the brane.
If one starts from the bulk terms
the model can be motivated by radiative generation
of kinetic terms on the brane due to self energy contributions
from brane modes \cite{DGP,DGS,DG,DDGV,DGKN}.

Action principles like (\ref{eq:actionDG}) will be denoted
as dimensionally hybrid action principles in the sequel.

In the Newtonian limit (\ref{eq:actionDG}) yields the static gravitational
potential of a mass $M$ on the brane as \cite{DGP}
\begin{eqnarray}\label{eq:Udgp}
U(\vek{r})&=&-\frac{M}{6\pi m_3^2 r}\left[
\cos\!\left(\frac{2m_4^3}{m_3^2}r\right)\right.
-
\frac{2}{\pi}\cos\!\left(\frac{2m_4^3}{m_3^2}r\right)
\mbox{Si}\!\left(\frac{2m_4^3}{m_3^2}r\right)  \\
 \nonumber
 &{}& \left.
+\frac{2}{\pi}\sin\!\left(\frac{2m_4^3}{m_3^2}r\right)
\mbox{ci}\!\left(\frac{2m_4^3}{m_3^2}r\right)
\right]
\end{eqnarray}
with the sine and cosine integrals
\[
\mbox{Si}(x)=\int_0^xd\xi\,\frac{\sin\xi}{\xi},
\qquad
\mbox{ci}(x)=-\int_x^\infty d\xi\,\frac{\cos\xi}{\xi}.
\]
$U(\vek{r})=U(\vek{r},0)$ is the gravitational potential at a point
$\vek{r}$ on the brane. The off brane gravitational
potential $U(\vek{r},x^\perp)$ can be expressed
in terms of special functions using
\[
U(\vek{p},p_\perp)=-\frac{4}{3}\frac{M}{(\vek{p}^2+p_\perp^2)
(2m_4^3+m_3^2|\vek{p}|)}.
\]
(\ref{eq:Udgp}) yields
four-dimensional gravity at short distances and five-dimensional gravity
at large distances
\[
r\ll\ell_{DGP}:
\]
\[
U(\vek{r})=-\frac{M}{6\pi m_3^2 r}
\left[1+\left(\gamma-\frac{2}{\pi}\right)\frac{r}{\ell_{DGP}}
+\frac{r}{\ell_{DGP}}\ln\!\left(\frac{r}{\ell_{DGP}}\right)
+\mathcal{O}\!\left(\frac{r^2}{\ell^2_{DGP}}\right)\right],
\]
\[
r\gg\ell_{DGP}:
\]
\[
U(\vek{r})=-\frac{M}{6\pi^2 m_4^3 r^2}
\left[1-2\frac{\ell^2_{DGP}}{r^2}
+ \mathcal{O}\!\left(\frac{\ell^4_{DGP}}{r^4}\right)\right],
\]
with the transition scale 
\begin{equation}\label{eq:ldgp}
\ell_{DGP}=\frac{m_3^2}{2m_4^3}.
\end{equation}

The presence of the higher-dimensional terms in the weak
field limit of the Einstein equation for $U=-h_{00}/2$ 
causes the modification in the numerical factor
between the reduced four-dimensional 
Planck mass $m_3$
and Newton's constant to $m_3=(6\pi G_N)^{-1/2}$ \cite{rd1}.

The Coulomb potential on the brane is
\begin{equation}\label{eq:Vdgs}
A^0(\vek{r})=\frac{q_3}{4\pi r}\left[
\cos\!\left(\frac{2q_3^2}{q_4^2}r\right)
-
\frac{2}{\pi}\cos\!\left(\frac{2q_3^2}{q_4^2}r\right)
\mbox{Si}\!\left(\frac{2q_3^2}{q_4^2}r\right)
+\frac{2}{\pi}\sin\!\left(\frac{2q_3^2}{q_4^2}r\right)
\mbox{ci}\!\left(\frac{2q_3^2}{q_4^2}r\right)
\right], 
\end{equation}
which corresponds to a transition scale
\begin{equation}\label{eq:ldgs}
\ell_{DGS}=\frac{q_4^2}{2q_3^2}
\end{equation}
between four-dimensional behavior at short distances and
 five-dimensional behavior at large distances \cite{DGS}.

A further attractive feature of the DGP model is that
it allows for an implementation of standard Friedmann cosmology
on the brane\footnote{See \cite{cedric,DDG,AM,DDG2,cedric2,DGS2} for
discussions of $\mathbb{Z}_2$ symmetric cosmology in the DGP
brane model, where Friedmann cosmology can be realized approximately.}
 \cite{rd1,CV,rd2}.

The motivation for the present work was twofold: On the one hand we were
seeking a corresponding model for fermions with both brane and bulk terms,
and on the other hand we wanted to understand the impact of a bulk mass term
for the gauge fields in this class of large extra dimension models.
Massless bulk gravitons can be acceptable due to their extremely weak 
coupling and the great difficulties of observing
gravitational waves, but for photons the time-dependent 
propagator following from (\ref{eq:actionDG})
would imply unacceptable radiation loss into the transverse dimension
from dynamical sources on the brane.

Dynamical binding mechanisms of matter and gravity to submanifolds as
a feature of solitonic solutions or as a consequence of couplings
to solitonic backgrounds \cite{history,cvetic1,cvetic2} has been discussed well 
before the recent formulation
of brane models in terms of dimensionally hybrid action principles.
One possibility
is to think of the models discussed here as effective descriptions
of underlying binding mechanisms e.g.\ in string inspired
brane models. Of course, this may not be the only possibility in which
actions like (\ref{eq:actionDG}) or (\ref{eq:actionA},\ref{eq:actionpsi})
below may arise.

We consider the impact of a bulk photon mass in Sec.\ \ref{sec:photon}
and bulk fermion masses in Sec.\ \ref{sec:fermion}.

\section{A photon mass in the bulk}\label{sec:photon}

With brane sources and a bulk mass term the action for photons on and off the
brane becomes\footnote{We assume that the vacuum solutions of the
underlying brane model allow for expansion around
a flat background, as is the case
e.g.\ in the DGP model \cite{DGP,rd1,GR,mk}.}
\begin{eqnarray}\label{eq:actionA}
S&=&\left.\int d^4x\left(
j^\mu A_\mu-\frac{1}{4}F_{\mu\nu}F^{\mu\nu}
\right)\right|_{x^\perp=0}\\
\nonumber
&&
+
\mathcal{M}\int d^4x\int dx^\perp
\left(
-\frac{1}{4}F_{MN}F^{MN}
-\frac{1}{2}M^2 A^M A_M\right).\nonumber
\end{eqnarray}
The source terms $j^\mu$ on the brane
are assumed to satisfy a conservation law
\begin{equation}\label{eq:dj}
\partial_\mu j^\mu=0.
\end{equation}

There are two mass scales in the problem: $\mathcal{M}$ determines
the relative weight of the bulk and brane contributions, and from
the results of \cite{DGS} we know that small $\mathcal{M}$ corresponds
to a large transition length $\ell_{DGS}$
to five-dimensional behavior of the static Coulomb potential.
However, for $M=0$
and time-dependent brane sources we would find radiation leaking into the bulk
also at length scales much smaller than $\ell_{DGS}$, and therefore
the bulk photon mass term has been included to suppress missing 
energy from radiation loss.

The equations of motion 
\[
\mathcal{M}\left(\partial_M F^{MN}-M^2 A^N\right)
+\delta(x^\perp)\eta^N{}_\nu\partial_\mu F^{\mu\nu}
=-\delta(x^\perp)\eta^N{}_\nu j^\nu
\]
split into
\begin{equation}\label{eq:divA}
\partial_M A^M=0
\end{equation}
and
\begin{equation}\label{eq:divF}
\mathcal{M}\left(\partial_M\partial^M A^N-M^2 A^N\right)
+\delta(x^\perp)\eta^N{}_\nu\left(\partial_\mu\partial^\mu A^\nu
+\partial^\nu\partial_\perp A^\perp\right)
=-\delta(x^\perp)\eta^N{}_\nu j^\nu.
\end{equation}

$A^\perp$ should be symmetric across the brane and
decouples from the sources. We set it to zero in the sequel.
To determine the potentials
\[
A_\mu(t,\vek{x},x^\perp)
=\int dt'\int d^3\vek{x}'\,G(t-t',\vek{x}-\vek{x}',x^\perp) j_\mu(t',\vek{x}')
\]
for the electromagnetic fields generated by the currents on the brane 
we have to solve
\begin{equation}\label{eq:greenA}
\mathcal{M}\left(\partial_M\partial^M-M^2\right)G(x,x^\perp)
+\delta(x^\perp)\partial_\mu\partial^\mu G(x,0)=-\delta^4(x)\delta(x^\perp).
\end{equation}
The Fourier {\it ansatz}
\[
G(x,x^\perp) = \frac{1}{(2\pi)^5}\int d^4p\, dp_\perp\,
G(p,p_\perp) 
\exp[\mathrm{i}(p\cdot x+p_\perp x^\perp)]
\]
yields
\[
\mathcal{M}\left(p^2+p_\perp^2+M^2\right)G(p,p_\perp)
+\frac{p^2}{2\pi}
\int_{-\infty}^\infty dp'_\perp\, G(p,p'_\perp)=1,
\]
which implies the following form for the Green's function
\[
G(p,p_\perp)=\frac{f(p)}{p^2+p_\perp^2+M^2}.
\]
With
\[
\frac{1}{2\pi}\mathcal{P}\!\!\int_{-\infty}^\infty\! dp'_\perp\,
\frac{1}{p^2+p_\perp^2+M^2}=\frac{\Theta(p^2+M^2)}{2\sqrt{p^2+M^2}}
\]
we then find for $p^2+M^2=\vek{p}^2+M^2-E^2>0$, i.e.\
below the threshold set by the bulk photon mass:
\begin{equation}\label{eq:GlowE1}
G(p,p_\perp)
=
\frac{2\sqrt{p^2+M^2}}{(2\mathcal{M}\sqrt{p^2+M^2}+p^2)(p^2+p_\perp^2+M^2)},
\end{equation}
\begin{equation}\label{eq:GlowE2}
G(p,x^\perp)=
\frac{\exp\!\left(-\sqrt{p^2+M^2}|x^\perp|\right)}{2\mathcal{M}\sqrt{p^2+M^2}+p^2},
\end{equation}
and above the threshold
\begin{equation}\label{eq:GhighE1}
G(p,p_\perp)=
\frac{1}{\mathcal{M}(p^2+p_\perp^2+M^2)},
\end{equation}
\begin{equation}\label{eq:GhighE2}
G(p,x^\perp) = 
-\frac{1}{2\mathcal{M}\sqrt{-p^2-M^2}} 
\sin\!\left(\sqrt{-p^2-M^2}|x^\perp|\right).
\end{equation}

 For $\mathcal{M}\to 0$ we find the ordinary 4-dimensional potential
$G(p)=G(p,x^\perp)|_{x^\perp=0}$ below the bulk photon threshold.
 For $\mathcal{M}\to\infty$ we find $G\to 0$. This is correct, since
$G$ describes the potentials generated from brane sources, and in the 
limit $\mathcal{M}\to\infty$ the electromagnetic fields decouple from the
brane sources. If we want to calculate the fields from bulk sources
residing on the brane we would have to rescale the source terms on the
right hand side of (\ref{eq:greenA}) by $\mathcal{M}$, which of course amounts
to a rescaling of $G$ by the same factor. The limit $\lim_{\mathcal{M}\to\infty}
\mathcal{M}G$ yields exactly the usual 5-dimensional potentials, as expected.

\section{Fermion masses in the bulk}\label{sec:fermion}

To formulate a dimensionally hybrid action principle for fermions
we can take e.g.\ 
$\gamma^\perp=\mathrm{i}\gamma_5=\gamma_0\gamma_1\gamma_2\gamma_3$.
The formulation of the action principle introduces again a bulk fermion
mass and an extra mass scale $\mathcal{M}$:
\begin{equation}\label{eq:actionpsi}
S=\left.\int d^4x\,\overline{\psi}\left(
\mathrm{i}\gamma^\mu
\partial_\mu-m\right)\psi\right|_{x^\perp=0}
+\mathcal{M}\!
\int\! d^4x\!
\int\! dx^\perp\,
\overline{\psi}\left(
\mathrm{i}\gamma^\mu
\partial_\mu+\mathrm{i}\gamma^\perp\partial_\perp-M\right)\psi.
\end{equation}
The corresponding equation for the free fermion propagator
for sources on the brane is
\begin{equation}\label{eq:greenpsi}
\mathcal{M}\left(
\mathrm{i}\gamma^\mu
\partial_\mu+\mathrm{i}\gamma^\perp\partial_\perp-M\right)S(x,x^\perp)
+\delta(x^\perp)
\left(
\mathrm{i}\gamma^\mu
\partial_\mu-m\right)S(x,0)=-\delta^4(x)\delta(x^\perp),
\end{equation}
and with
\[
S(x,x^\perp) = 
\frac{1}{(2\pi)^5}\int d^4p\, dp_\perp\,
S(p,p_\perp)  
\exp[\mathrm{i}(p\cdot x+p_\perp x^\perp)]
\]
we find
\begin{equation}\label{eq:condS}
  \mathcal{M}\left(
\gamma^\mu p_\mu+\gamma^\perp p_\perp+M\right)S(p,p_\perp) 
  +\frac{\gamma^\mu p_\mu+m}{2\pi}
\int_{-\infty}^\infty dp'_\perp\, S(p,p'_\perp)=1.
\end{equation}
This determines the $p_\perp$-dependence of the propagator
\[
S(p,p_\perp)=
\frac{M-\gamma^\mu p_\mu-\gamma^\perp p_\perp}{M^2+p^2+p_\perp^2}
s(p),
\]
and $s(p)$ is then determined from (\ref{eq:condS}) 
by taking into account
\[
\frac{1}{2\pi}\mathcal{P}\!\!\int_{-\infty}^\infty\! dp_\perp\,
\frac{M-\gamma^\mu p_\mu-\gamma^\perp p_\perp}{M^2+p^2+p_\perp^2}
=\frac{M-\gamma^\mu p_\mu}{2\sqrt{p^2+M^2}}\Theta(p^2+M^2).
\]
At low energies $p^2+M^2>0$ this yields again exponential damping
in the transverse direction:
\begin{equation}\label{eq:SlowE1}
S(p,p_\perp)=
2\sqrt{p^2+M^2}
\frac{M-\gamma^\mu p_\mu-\gamma^\perp p_\perp}{M^2+p^2+p_\perp^2}
\end{equation}
\[
\times
\frac{
2\mathcal{M}\sqrt{p^2+M^2}+mM+p^2+(m-M)\gamma^\mu p_\mu
}
{
4\mathcal{M}^2(p^2+M^2)+4\mathcal{M}\sqrt{p^2+M^2}(mM+p^2)+p^4
+(m^2+M^2)p^2+m^2M^2,
}
\]
\begin{equation}\label{eq:SlowE2}
S(p,x^\perp)=
\left(
M-\gamma^\mu p_\mu-\mathrm{i}\gamma^\perp\sqrt{p^2+M^2}\mathrm{sign}(x^\perp)
\right)
\exp\!\left(-\sqrt{p^2+M^2}|x^\perp|\right)
\end{equation}
\[
\times
\frac{
2\mathcal{M}\sqrt{p^2+M^2}+mM+p^2+(m-M)\gamma^\mu p_\mu
}
{
4\mathcal{M}^2(p^2+M^2)+4\mathcal{M}\sqrt{p^2+M^2}(mM+p^2)+p^4
+(m^2+M^2)p^2+m^2M^2
},
\]
while above the threshold for bulk fermions we have
\begin{equation}\label{eq:ShighE1}
S(p,p_\perp)=
\frac{1}{\mathcal{M}}
\frac{M-\gamma^\mu p_\mu-\gamma^\perp p_\perp}{M^2+p^2+p_\perp^2},
\end{equation}
\begin{eqnarray}\label{eq:ShighE2}
S(p,x^\perp)&=&
-\frac{1}{2\mathcal{M}}\left[
\frac{M-\gamma^\mu p_\mu}{\sqrt{-p^2-M^2}}
\sin\!\left(\sqrt{-p^2-M^2}|x^\perp|\right)
\right. \\
\nonumber
&&\left.
+\mathrm{i}\gamma^\perp\mathrm{sign}(x^\perp)
\cos\!\left(\sqrt{-p^2-M^2}|x^\perp|\right)
\right].
\end{eqnarray}

The fermion propagators have the usual hermiticity
properties
\[
\gamma^0 S^+(p,p_\perp)\gamma^0=S(p,p_\perp)
\]
\[
\gamma^0 S^+(p,x^\perp)\gamma^0=S(p,-x^\perp),
\]
and the expected 4-dimensional and 5-dimensional limiting
behavior:
$\lim_{\mathcal{M}\to\infty} \mathcal{M}S$ is the
usual 5-dimensional fermion propagator, and 
below the bulk fermion threshold we have
\[
\lim_{\mathcal{M}\to 0}S(p)=
\left.\lim_{\mathcal{M}\to 0}S(p,x^\perp)\right|_{x^\perp=0}
=\frac{m-\gamma^\mu p_\mu}{m^2+p^2}.
\]

\section{Conclusion}\label{sec:conc}

The dimensionally hybrid action principles (\ref{eq:actionA},\ref{eq:actionpsi})
yield propagators which
interpolate between genuine four-dimensional propagators at small distances 
$\ll\mathcal{M}^{-1}$ and
five-dimensional propagators at large distances $\gg\mathcal{M}^{-1}$.
However,
suppression of radiation into the bulk requires bulk masses $M\gg\mathcal{M}$,
and as a consequence of that the four-dimensional behavior actually persists
only in an energy range $\mathcal{M}\ll E\ll M$. Furthermore, the bulk mass
shifts the pole of the photon propagator $G(p)=G(p,x^\perp)|_{x^\perp}=0$
 also in the four-dimensional regime to
\[
m_\gamma^2=2\mathcal{M}\left(\sqrt{\mathcal{M}^2+M^2}-\mathcal{M}\right)
\approx 2\mathcal{M}M.
\]
That this corresponds to the generation of a photon mass in four dimensions
may not be as obvious as in a genuine four-dimensional
propagator, since the pole appears as a simple algebraic pole only
in an expansion of the denominator in $G(p)$ around $m_\gamma^2$,
and also the integration to $G(x)$ is not straightforward.

However, that $m_\gamma$ is indeed an effective photon mass on the brane
can also be inferred from the free equation
\[
\mathcal{M}\left(\partial_M\partial^M-M^2\right)A_\nu(x,x^\perp)
+\delta(x^\perp)\partial_\mu\partial^\mu A_\nu(x,0)=0,
\]
which yields
\[
A_\mu(p,p_\perp)\propto
\frac{1}{p_\perp^2+\left(\sqrt{\mathcal{M}^2+M^2}-\mathcal{M}\right)^2}
\delta\left(p^2
+2\mathcal{M}\sqrt{\mathcal{M}^2+M^2}-2\mathcal{M}^2\right).
\]

The corresponding mass shift for massive fields  (\ref{eq:SlowE2}) 
is of much less significance, since generically it will provide only a
small shift of $m$: In leading order in $\mathcal{M}$ the shift of a
mass $m<M$ is
\[
\delta m^2=4m\mathcal{M}\sqrt{\frac{M-m}{M+m}}.
\]

The current upper bound on a photon mass is \cite{lakes,PDG}
\[
m_\gamma< 2\times 10^{-16}\,\mathrm{eV}.
\]
No significant missing energy events have been found in accelerator
searches up to
$\sqrt{s}=1.8\,$TeV. Therefore
bulk photon masses in the class of brane models 
discussed here must certainly exceed $M>1\,$TeV. This translates into
a bound on the transition scale to five-dimensional behavior
\[
(2\mathcal{M})^{-1}=\frac{M}{m_\gamma^2}>1.7\times 10^{11}\,\mathrm{Gpc},
\]
which exceeds the size of our Hubble horizon by a factor $\approx 2\times
10^{10}$:
In those brane models where photon localization is
dynamical and can effectively be described by a bulk mass term, laboratory
constraints on brane and bulk photon masses are more restrictive than 
constraints from large scale observations of four-dimensional
distance laws.\\[1ex]
{\bf Acknowledgement:} This work was supported in part by NSERC Canada.

\end{document}